\newcommand\apjcls{1}
\newcommand\aastexcls{2}
\newcommand\othercls{3}

\newcommand\papercls{\aastexcls}
\documentclass[tighten, times,twocolumn]{aastex62}  

\if\papercls \apjcls
\usepackage{apjfonts}
\else\if\papercls \othercls
\usepackage{epsfig}
\usepackage{margin}
\usepackage{times}
\fi\fi
\usepackage{ifthen}
\usepackage{natbib}
\usepackage{bm}
\usepackage{amssymb, amsmath}
\usepackage{appendix}
\usepackage{etoolbox}
\usepackage[T1]{fontenc}
\usepackage{paralist}
\usepackage{newtxtext,newtxmath}
\if\papercls \apjcls
\newcommand\aas{\ref@jnl{AAS Meeting Abstracts}}
\newcommand\dps{\ref@jnl{AAS/DPS Meeting Abstracts}}
\newcommand\maps{\ref@jnl{MAPS}}
\else\if\papercls \othercls
\usepackage{astjnlabbrev-jh}
\fi\fi

\usepackage{lipsum}  
\if\papercls \aastexcls
\hypersetup{citecolor=blue, 
            linkcolor=blue, 
            menucolor=blue, 
            urlcolor=blue}  
\else
\usepackage[
bookmarks=true,           
bookmarksnumbered=true,   
colorlinks=true,          
citecolor=blue,           
linkcolor=blue,           
menucolor=blue,           
urlcolor=blue,            
linkbordercolor={0 0 1},  
pdfborder={0 0 1},
frenchlinks=true]{hyperref}
\fi

\if\papercls \othercls

\else

\fi

\providecommand{\adsurl}[1]{\href{#1}{ADS}}

\makeatletter
\patchcmd{\NAT@citex}
  {\@citea\NAT@hyper@{%
     \NAT@nmfmt{\NAT@nm}%
     \hyper@natlinkbreak{\NAT@aysep\NAT@spacechar}{\@citeb\@extra@b@citeb}%
     \NAT@date}}
  {\@citea\NAT@nmfmt{\NAT@nm}%
   \NAT@aysep\NAT@spacechar\NAT@hyper@{\NAT@date}}{}{}

\patchcmd{\NAT@citex}
  {\@citea\NAT@hyper@{%
     \NAT@nmfmt{\NAT@nm}%
     \hyper@natlinkbreak{\NAT@spacechar\NAT@@open\if*#1*\else#1\NAT@spacechar\fi}%
       {\@citeb\@extra@b@citeb}%
     \NAT@date}}
  {\@citea\NAT@nmfmt{\NAT@nm}%
   \NAT@spacechar\NAT@@open\if*#1*\else#1\NAT@spacechar\fi\NAT@hyper@{\NAT@date}}
  {}{}
\makeatother

\makeatletter
\DeclareRobustCommand{\lowcase}[1]{\@lowcase#1\@nil}
\def\@lowcase#1\@nil{\if\relax#1\relax\else\MakeLowercase{#1}\fi}
\makeatother

\DeclareSymbolFont{UPM}{U}{eur}{m}{n}
\DeclareMathSymbol{\umu}{0}{UPM}{"16}
\let\oldumu=\umu
\renewcommand\umu{\ifmmode\oldumu\else\math{\oldumu}\fi}

\if\papercls \othercls

\else

\fi

\let\oldsim=\sim
\renewcommand\sim{\ifmmode\oldsim\else\math{\oldsim}\fi}
\let\oldpm=\pm
\renewcommand\pm{\ifmmode\oldpm\else\math{\oldpm}\fi}
\newcommand\by{\ifmmode\times\else\math{\times}\fi}

\newbox{\wdbox}
\renewcommand\c{\setbox\wdbox=\hbox{,}\hspace{\wd\wdbox}}
\renewcommand\i{\setbox\wdbox=\hbox{i}\hspace{\wd\wdbox}}

\newcount\timect
\newcount\hourct
\newcount\minct
\newcommand\now{\timect=\time \divide\timect by 60
         \hourct=\timect Cltiply\hourct by 60
         \minct=\time \advance\minct by -\hourct
         \number\timect:\ifnum \minct < 10 0\fi\number\minct}

\catcode`@=11

\newcommand\comment[1]{}

\newcommand\commenton{\catcode`\%=14}

\renewcommand\math[1]{$#1$}
\newcommand\mathshifton{\catcode`\$=3}

\let\atab=&
\newcommand\atabon{\catcode`\&=4}

\let\oldmsp=\sp
\let\oldmsb=\sb
\def\sp#1{\ifmmode
           \oldmsp{#1}%
         \else\strut\raise.85ex\hbox{\scriptsize #1}\fi}
\def\sb#1{\ifmmode
           \oldmsb{#1}%
         \else\strut\raise-.54ex\hbox{\scriptsize #1}\fi}
\newbox\@sp
\newbox\@sb
\def\sbp#1#2{\ifmmode%
           \oldmsb{#1}\oldmsp{#2}%
         \else
           \setbox\@sb=\hbox{\sb{#1}}%
           \setbox\@sp=\hbox{\sp{#2}}%
           \rlap{\copy\@sb}\copy\@sp
           \ifdim \wd\@sb >\wd\@sp
             \hskip -\wd\@sp \hskip \wd\@sb
           \fi
        \fi}
\def\msp#1{\ifmmode
           \oldmsp{#1}
         \else \math{\oldmsp{#1}}\fi}
\def\msb#1{\ifmmode
           \oldmsb{#1}
         \else \math{\oldmsb{#1}}\fi}

\def\supon{\catcode`\^=7}

\def\subon{\catcode`\_=8}

\def\supsubon{\supon \subon}

\newcommand\actcharon{\catcode`\~=13}

\newcommand\paramon{\catcode`\#=6}

\comment{And now to turn us totally on and off...}

\newcommand\reservedcharson{ \commenton  \mathshifton  \atabon  \supsubon 
                             \actcharon  \paramon}

\catcode`@=12
\reservedcharson

\if\papercls \apjcls

\else

\fi

\newcommand\chisq{\ifmmode{\chi\sp{2}}\else\math{\chi\sp{2}}\fi}
\newcommand\redchisq{\ifmmode{ \chi\sp{2}\sb{\rm red}}
                    \else\math{\chi\sp{2}\sb{\rm red}}\fi}
\newcommand\Teq{\ifmmode{T\sb{\rm eq}}\else$T$\sb{eq}\fi}
\newcommand\mjup{\ifmmode{M\sb{\rm Jup}}\else$M$\sb{Jup}\fi}
\newcommand\rjup{\ifmmode{R\sb{\rm Jup}}\else$R$\sb{Jup}\fi}
\newcommand\msun{\ifmmode{M\sb{\odot}}\else$M\sb{\odot}$\fi}
\newcommand\rsun{\ifmmode{R\sb{\odot}}\else$R\sb{\odot}$\fi}
\newcommand\mearth{\ifmmode{M\sb{\oplus}}\else$M\sb{\oplus}$\fi}
\newcommand\rearth{\ifmmode{R\sb{\oplus}}\else$R\sb{\oplus}$\fi}

\renewcommand{\bm}[1]{{\mbox{{\boldmath$#1$}}}}	

\newcommand{\grad}{\bm{\nabla}}

\shorttitle{Spin-up in superfluid neutron stars}
\shortauthors{Fuentes and Graber}

\begin{document}

\title{Superfluid Spin-up: 3D Simulations of Post-Glitch Dynamics in Neutron Star Cores}

\author{J. R. Fuentes}
\affiliation{\rm Department of Applied Mathematics, University of Colorado Boulder, Boulder, CO 80309-0526, USA}

\author{Vanessa Graber}
\affiliation{\rm  Department of Physics, Astronomy and Mathematics, University of Hertfordshire, Hatfield, Hertfordshire, AL10 9AB, United Kingdom}

\begin{abstract}

Neutron stars show a steady decrease in their rotational frequency, occasionally interrupted by sudden spin-up events called glitches. The dynamics of a neutron star after a glitch involve the transfer of angular momentum from the crust (where the glitch is presumed to originate) to the liquid core, causing the core to spin up. The crust-core coupling, which determines how quickly this spin-up proceeds, can be achieved through various physical processes, including Ekman pumping, superfluid vortex-mediated mutual friction, and magnetic fields. Although the complex nature of these mechanisms has made it difficult to study their combined effects, analytical estimations for individual processes reveal that spin-up timescales vary according to the relative strength of Coriolis, viscous, and mutual friction forces, as well as the magnetic field.
However, experimental and numerical validations of those analytical predictions are limited. In this paper, we focus on viscous effects and mutual friction. We conduct non-linear hydrodynamical simulations of the spin-up problem in a two-component fluid by solving the incompressible Hall–Vinen–Bekarevich–Khalatnikov (HVBK) equations in the full sphere (i.e., including $r=0$) for the first time. We find that the viscous (normal) component accelerates due to Ekman pumping, although the mutual friction coupling to the superfluid component alters the spin-up dynamics compared to the single-fluid scenario. Close to the sphere's surface, the response of the superfluid is accurately described by the mutual friction timescale irrespective of its coupling strength with the normal component. However, as we move deeper into the sphere, the superfluid accelerates on different timescales due to the slow viscous spin-up of the internal normal fluid layers. We discuss potential implications for neutron stars and requirements for future work to build more realistic models.

\end{abstract}

\keywords{
                dense matter — hydrodynamics — stars: interiors — stars: neutron — stars: rotation}

\section{Introduction}
Neutron stars are the densest objects in the universe and directly observable across the electromagnetic spectrum, e.g., as radio pulsars, gamma-ray pulsars or as X-ray binaries. Their rotation rates can be measured with great precision, revealing a remarkably stable spin-down trend that is traditionally attributed to the loss of rotational energy caused by the emission of electromagnetic radiation \citep{Ryba_and_Taylor_1991a,Ryba_and_Taylor_1991b,Kaspi_1994}. One of the most exciting aspects of neutron stars is that in terms of their high densities, these compact objects are cold and thus quantum mechanics strongly influences their interiors. In particular, the neutrons and protons in the fluid core of the star, as well as the neutrons permeating the inner part of the solid crust, form Cooper pairs. This allows them to condense into a BCS-type superfluid state (or, in the case of the protons, a superconducting state) which impacts on the large-scale stellar dynamics \citep[see, e.g.,][]{Chamel2017_sup,Haskell2018}. Specifically, macroscopic superfluidity affects the neutron star rotation. In roughly 15\% of isolated neutron stars with characteristic ages younger than $10^{7} \, {\rm yr}$, the secular spin-down trend is interrupted by \emph{glitches}, i.e., instances where the rotation frequency of the star suddenly increases \citep[see, e.g.,][]{Espinoza_2011,Yu2013,Fuentes_2017,Antonopoulou_2022,Millhouse2022}. These glitches have fractional sizes of $\Delta \Omega / \Omega \sim 10^{-10} - 10^{-5}$ (where $\Omega$ is the star's rotation frequency and $\Delta \Omega$ the glitch size) and have been associated with the macroscopic manifestation of superfluidity inside neutron stars. Although the global and mesoscopic flow patterns in the interior of a neutron star are unknown, laboratory experiments using superfluid helium and ultracold gases have shown that such quantum condensates rotate by forming an array of quantized vortices whose areal density determines the spin rate of the whole superfluid body \citep[e.g.,][]{Vinen1961,Hadzibabic2006,Schweikhard2007,Bewley2008}. These findings have motivated the development of several vortex-based models for the origin of glitches \citep[see, e.g, the review by][]{Haskell_and_Melatos_2015}. 

In the standard superfluid glitch model \citep[e.g.,][]{Anderson_1975}, the observed spin-up is explained as follows: While the outer crust (and all the stellar components that are tightly coupled to it) slows down owing to electromagnetic energy losses, the neutron superfluid does not. This is due to the fact that bulk superfluid spin-down corresponds to a decrease in areal vortex density, achieved by vortices moving radially outwards and annihilating at the outer boundary of the superfluid region. However, in the inner crust, vortices can become pinned to the lattice nuclei, effectively preventing the neutron superfluid from spinning down and building up an angular momentum reservoir over time. Once the difference between the stellar spin rate and that of the neutron superfluid in the inner crust exceeds a critical threshold, vortices unpin catastrophically and transfer their excess angular momentum to the crust. Angular momentum conservation is thus responsible for the increase in crustal rotation rate, producing what we observe as a glitch.

However, during a glitch, the redistribution of angular momentum is not only limited to the crust. In fact, angular momentum is also transferred to the multi-component fluid core, causing the innermost neutron star regions to spin up as well \citep{Graber2018,Pizzochero2020,Gugercinoglu2020}. This process is extremely important, as the resulting fluid motions control the relaxation and recovery phase of the neutron star after a glitch, and may excite oscillation modes and non-axisymmetric perturbations, triggering gravitational waves (GWs) \citep{Bennett2010, Prix2011, Passamonti2011, Lasky2015, Yim2020, Yim2024}. Recent calculations have suggested that GWs triggered by glitches might be marginally detectable with Advanced LIGO at design sensitivity, while they are likely detectable for third-generation detectors such as the Einstein Telescope \citep{Moragues2023}. Therefore, understanding the global dynamics of multi-component fluids and core-crust coupling following a glitch is of great relevance.

Before discussing previous research on the spin-up of superfluids, it is helpful to comment on the classic spin-up problem of a (single-component) viscous fluid first. Imagine a container filled with water that is rotating uniformly. If the container is suddenly accelerated, the interior fluid is observed to spin up on a timescale that is much shorter than what it would be due to viscous diffusion. This phenomenon, known as Ekman pumping \citep[e.g.,][]{Greenspan_and_Howard_1963,Benton_and_Clark_1974}, proceeds as follows: A viscous boundary layer forms on a timescale comparable to the rotation period of the container. While the interior fluid initially continues to rotate at the original container speed, the fluid in the boundary layer rotates at the new (and larger) container speed. Low angular momentum fluid entering the boundary layer from the interior is subsequently replaced by fluid with greater angular momentum (convected inward to conserve mass) in such a way that the interior fluid achieves solid-body rotation at the new container's speed.

Superfluids are more complex than classical fluids. Based on initial studies of superfluid helium, their hydrodynamics (valid in the limit where lengthscales in the flow are longer than the average separation between vortex lines) are typically described using two separate components \citep{Tisza1938, Landau1941, Barenghi1983}. The first component is a pure superfluid of zero viscosity, and the second component is a ``normal fluid'' (the thermal excitations of the superfluid) which has a finite viscosity. The relative fraction of these two components depends strongly on temperature $T$ as the normal component vanishes as we approach $T \to 0$ while the superfluid fraction disappears at the critical temperature $T_{\rm c}$. Both components are coupled by a friction force that arises from the scattering of thermal excitations off the normal cores of the vortex lines (an effect referred to as mutual friction).

\cite{Tsakadze_and_Tsakadze_1973,Tsakadze_and_Tsakadze_1975,Tsakadze_and_Tsakadze_1980} investigated the spin-up (and spin-down) of a vessel filled with He II (superfluid $^4$He that exists at temperatures below the ``lambda point'', $T_{\lambda} = 2.17~\mathrm{K}$) for different temperatures, rotation rates, and container spin changes. Unlike for the normal fluid, they found that the spin-up of a superfluid occurs on much longer timescales that have a tendency to decrease with increasing temperature. A detailed theoretical interpretation of Tsakadzes' results was given by \cite{Reisenegger_1993}, who studied the laminar spin-up of He II confined between two parallel planes \citep[see also][for a treament of the same problem using different geometries]{vanEysden_2011,van_Eysden_2013}. At low temperatures, the spin-up process is dominated by the mutual friction between the superfluid vortex lines and a viscous boundary layer coupled to the container. This friction produces an Ekman-like circulation that transports the vortex lines and angular momentum inwards. In contrast, at higher temperatures, much closer to $T_\lambda$, the spin-up mechanism is dominated by Ekman pumping in the normal fluid. The superfluid also spins up because it is coupled to the normal fluid by the mutual friction force.

A comprehensive numerical study of two-component spin-up in the context of superfluid neutron stars was conducted in a series of papers by \cite{Peralta2005,Peralta_2006,Peralta2008}, focusing on understanding the global flow pattern and a range of superfluid instabilities in differentially rotating spherical shells. Note that in the context of neutron stars, the two components do not correspond to an inviscid superfluid component and the finite-temperature condensate of excitations as in He II. Instead, mature neutron stars, whose internal temperatures fall well below the critical temperatures for the onset of neutron and proton superfluidity \citep[e.g.,][]{Ho2015b}, are typically decomposed into a neutron superfluid and a normal conducting particle conglomerate that combines the electrons and protons; i.e., superconductivity is ignored \citep[e.g.,][]{Andersson2006, Glampedakis2011}. Using the same numerical model as \cite{Peralta2005}, \cite{Howitt2016} investigated the recovery phase after an induced glitch. They found that the crust's angular velocity can evolve in different ways, depending on the strength of the mutual friction force, and the location where the glitch is originated.

Besides the investigations mentioned above, the specific problem of the spin-up of a two-component superfluid in non-plane-parallel geometries remains relatively unstudied, even experimentally. Furthermore, previous direct numerical simulations in three dimensions have been limited to simulating shells, i.e., the coordinate singularity at $r=0$ is avoided by making a ``cut-out'' in the center of the sphere, forcing the use of additional boundary conditions at the inner surface of the shell. While the full impact of a core cut-out on the global solution is unclear, it certainly affects the flow morphology and spin-up timescales. Determining this effect requires extensive computational calibration, experimentation, and simulations on the full sphere. Using new techniques in the open-source pseudospectral framework Dedalus \citep{Vasil2019,Lecoanet2019,Burns2020}, we have performed the first numerical simulations of spherical spin-up that includes $r=0$ without the need for a cut-out.

In Section~\ref{sec:hydro}, we describe the set of equations that model the hydrodynamics of a two-component superfluid, and discuss the relevant timescales involved in the spin-up problem. In Section~\ref{sec:exp_methods}, we present the physical model and the numerical code used to perform the simulations. In Section~\ref{sec:results}, we show our main results. We first discuss the spin-up of a single-component fluid in spherical geometry and then address the two-component setup, focusing specifically on the timescale at which the interior flow spins up. Finally, we conclude in Section~\ref{sec:discussion} with a discussion.

\section{Multifluid hydrodynamics} 
\label{sec:hydro}

\subsection{Equations of motion}

We model the neutron star as a system of two coupled fluids: a superfluid of neutrons of mass density $\rho_s$ and velocity $\bm{u}_s$, and a normal (viscous) fluid of protons and electrons of density $\rho_n$ and velocity $\bm{u}_n$\footnote{We assume that protons and electrons are locked together due to electromagnetic coupling \citep{Mendell1991}.}. While the electrons remain in a normal state, protons are expected to undergo a transition into a superconducting state soon after the neutron star is born \citep{Sauls1989}. However, we ignore this effect to keep the analysis as simple as possible. For the same reason, we also assume that both fluids are incompressible (i.e., $\rho_s$ and $\rho_n$ are constants) and have zero temperature. Finally, under real conditions inside neutron stars, protons and neutrons are also coupled via the so-called entrainment effect, in which the momentum of one species is partly carried along by the other species \citep{Mendell1991,Sedrakian1995}. Neglecting the entrainment effect is likely a reasonable approximation in the star’s outer core \citep[see, e.g.,][]{Carter_Chamel_Haensel_2006}, while entrainment significantly impacts on the stellar crust \citep[e.g.,][]{Andersson2012, Chamel2013, Chamel2017, Sauls2020}. Since we are interested in the spin-up of the bulk interior fluid, we neglect entrainment in what follows.

Under the model assumptions described above, the fluid equations that capture the hydrodynamics of the system are similar to the Hall–Vinen–Bekarevich–Khalatnikov (HVBK) equations used to study laboratory superfluid helium \citep{HV1956a,HV1956b,Barenghi_and_Jones_1988}:

\begin{align}
&\dfrac{\partial\bm{u}_n}{\partial t} + \bm{u}_n\cdot \grad \bm{u}_n = - \grad \tilde{\mu}_n + \nu\grad^2 \bm{u}_n + \dfrac{\bm{F}_{\rm{MF}}}{\rho_p}~, \label{eq:u_n}\\
&\dfrac{\partial\bm{u}_s}{\partial t} + \bm{u}_s\cdot \grad \bm{u}_s  = - \grad \tilde{\mu}_s - \dfrac{\bm{F}_{\rm{MF}}}{\rho_s}~,\label{eq:u_s}\\
&\grad\cdot\bm{u}_n = 0 = \grad\cdot\bm{u}_s~, \label{eq:mass_conservation}
\end{align}
where $\rho = \rho_s + \rho_n$ is the total density, $\nu$ is the kinematic viscosity of the normal fluid, and $\tilde{\mu}_{s,n}$ denotes the chemical potential of the superfluid (subscript $s$) and normal component (subscript $n$), respectively. The pressure gradients and gravitational acceleration are taken to be constant and absorbed within the chemical potential terms. While this is not ideal in the full sphere, we adopt this assumption to reduce the complexity of our problem and compare our results with previous work \citep{van_Eysden_2013}. The mutual friction force $\bm{F}_{\mathrm{MF}}$ (responsible for coupling both fluids) arises from the interaction between the quantized superfluid vortex lines and the normal fluid. We present and discuss the mutual friction force in Section~\ref{sec:mf}.

\subsection{Coupling due to mutual friction}
\label{sec:mf}

Several mechanisms have been suggested to couple the superfluid and the normal particle conglomerate in the neutron star interior. For example, both components are connected via the dissipative scattering of electrons off of the vortex magnetic field \citep{Sauls1982, Alpar1984, Andersson_2006}. Another possibility is the magnetic interaction between vortices and quantized fluxtubes provided that protons form a type-II superconductor \cite[see, e.g.,][]{Ruderman1998,Link2003,Sidery2009,Sourie2020}. In the following, we focus on the former mechanisms due to the similarities of the underlying prescription with laboratory superfluid helium. In particular, the resulting mutual friction force can be derived by balancing the forces acting on each individual vortex, namely the Magnus force exerted by the neutrons and the resistive force exerted by the scattered electrons. The mutual friction force further depends on the configuration of the vortices \citep{Peralta2005}. Vortex lines have a tendency to resist bending, which results in a tension force that can be included in Equation~\eqref{eq:u_s}, and affects the form of $\bm{F}_{\mathrm{MF}}$ \citep{Andronikashvili1966,Hills_and_Roberts_1977}. For simplicity, we ignore vortex tension and vortex tangles triggered by any source of superfluid turbulence \citep{Gorter1949,Glaberson1974,Andersson2007}, and adopt the form that characterizes a (locally) straight array of vortices. In this approximation, the mutual friction force is given by

\begin{equation}\label{eq:mutual_friction}
\bm{F}_{\rm{MF}} = \rho_s\left[\mathcal{B}\left(\bm{\hat{\omega}}_s\times \left(\bm{\omega}_s \times \bm{u}_{sn}\right)\right) + \mathcal{B}'\left(\bm{\omega}_s \times \bm{u}_{sn}\right)\right]~,
\end{equation}
where $\bm{u}_{sn} = \bm{u}_s - \bm{u}_n$, $\bm{\omega}_s = \grad \times \bm{u}_s$, and $\mathcal{B}$ and $\mathcal{B}'$ are dimensionless coefficients related to the drag force experienced by individual neutron vortices. Dynamically, the term proportional to $\mathcal{B}$ is dissipative, whereas the second term proportional to $\mathcal{B}'$ is conservative \citep{Andersson_2006}. Calculations of mutual friction from electron scattering in neutron stars suggests $\mathcal{B}'\approx \mathcal{B}^2$, with $\mathcal{B}\sim 4\times 10^{-4}$ \citep{Mendell1991,Andersson_2006}.

We can get an estimate of the crust-core coupling timescale by calculating the body-averaged torque due to mutual friction \citep[e.g.,][]{Antonelli2022}. For simplicity, let us assume that the neutron superfluid and the normal component rotate uniformly with angular velocities $\bm{\Omega}_s$ and $\bm{\Omega}_n$ around the $z$-axis, respectively, so that the corresponding fluid velocities at a given position $\bm{r}$ are $\bm{u}_s = \bm{\Omega}_{s}\times \bm{r}$, and $\bm{u}_n = \bm{\Omega}_{n}\times \bm{r}$. Using spherical coordinates ($r,\theta,\phi$), the mutual friction force reads

\begin{equation}
\bm{F}_{\rm{MF}} = -2\rho_s\Omega_s \Delta\Omega_{sn} r \left(\mathcal{B}\sin\theta\bm{\hat{e}_{\phi}} + \mathcal{B}'\bm{\hat{e}_r} - \mathcal{B}'\cos\theta\bm{\hat{e}_z} \right)~,
\end{equation}
where $\Delta \Omega_{sn} = \Omega_s - \Omega_n$. We can then determine the cross product as

\begin{equation}\label{eq:rcrossF}
\bm{r}\times \bm{F}_{\rm{MF}} = 2\rho_s\Omega_s \Delta\Omega_{sn} r^2 \sin\theta \left(\mathcal{B}\bm{\hat{e}_{\theta}} - \mathcal{B}'\cos\theta\bm{\hat{e}_{\phi}}\right)~.
\end{equation}
The torque due to mutual friction is obtained by volume integrating Equation~\eqref{eq:rcrossF}. Note that since $\bm{\hat{e}_{\phi}} = -\sin\phi\bm{\hat{e}_x} + \cos\phi\bm{\hat{e}_y}$, the component $\propto \mathcal{B}'$ vanishes when performing the integration over $\phi$. Similarly, since $\bm{\hat{e}_{\theta}} = \cos\theta(\cos\phi\bm{\hat{e}_x} + \sin\phi\bm{\hat{e}_y}) - \sin\theta\bm{\hat{e}_z}$, only the $\bm{\hat{e}_z}$ component of the torque survives. Therefore:

\begin{align}\label{eq:torque}
\nonumber\bm{N}_{\rm{MF}} &= \int_V \bm{r}\times \bm{F}_{\rm{MF}}~ {\rm d} V~,\\
\nonumber &= -2\Omega_s \mathcal{B} \Delta\Omega_{sn} \int_V \rho_s r^2 \sin^2\theta~ {\rm d}V~\bm{\hat{e}_z}~,\\
&=-2\Omega_s \mathcal{B} \Delta\Omega_{sn} I_s~\bm{\hat{e}_z}~,
\end{align}
where we identified the integral in the second line as the moment of inertia of the superfluid of neutrons $I_s$ (recall that in spherical coordinates, the perpendicular distance to a point at $\bm{r}$ from the axis of rotation is $r_\perp = r\sin\theta$). Note that (as we would expect) the torque only depends on the dissipative part of the mutual friction and not on the conservative part, i.e., the contribution from $\mathcal{B}'$ is zero. Finally, since $N_{\mathrm{MF}} \sim \Delta L/\tau_{\rm{MF}}$, where $\Delta L \sim I_s \Delta\Omega_{sn}$ is the change in the angular momentum due to the applied torque, from Equation~\eqref{eq:torque}, we extract the the characteristic mutual friction spin-up timescale

\begin{equation}\label{eq:t_mf}
\tau_{\mathrm{MF}} \sim \dfrac{1}{2\Omega_s \mathcal{B}} \sim 80~{\mathrm s} \left(\dfrac{P_{\rm rot}}{0.1~\rm s}\right)\left(\dfrac{\mathcal{B}}{10^{-4}}\right)^{-1}~.
\end{equation}
As expected, the larger the friction coefficient $\mathcal{B}$, the shorter the coupling timescale.

For comparison, the Ekman time of the normal component is 

\begin{equation}\label{eq:Ek}
\tau_{\rm Ek} \sim \dfrac{1}{\sqrt{\mathrm{Ek}}\Omega_n}~,
\end{equation}
where $\mathrm{Ek} = \nu/2\Omega_n R^2$ is the Ekman number (assuming the characteristic length to be the radius $R$ of the star). For electron-electron scattering in the outer core of a neutron star, the viscosity is given by 

\begin{equation}
\nu \sim 10^5~\mathrm{cm^2~s^{-1}}\left(\dfrac{x_n}{0.05}\right)^{3/2}\left(\dfrac{\rho_n}{\mathrm{10^{14}~g~cm^{-3}}}\right)^{1/2}\left(\dfrac{T}{10^8~\mathrm{K}}\right)^{-2}~.
\end{equation}
\citep{Flowers1979,Cutler1987,Andersson2005}, where $x_n=\rho_n/\rho$ is the mass fraction of the normal component (which is equivalent to the proton mass fraction, given that in the interiors of neutron stars, the mass density of the electrons is negligible in comparison to that of the protons), and $T$ is the temperature. Then, for typical conditions in neutron stars, we estimate the Ekman number as

\begin{align} \label{eq:ek_n}
    \nonumber\mathrm{Ek} \sim 10^{-9}&\left(\dfrac{x_n}{0.05}\right)^{3/2}  \left(\dfrac{\rho_n}{10^{14}~\mathrm{g~cm^{-3}}}\right)^{1/2}  \left(\dfrac{P_{\mathrm{rot}}}{0.1~\mathrm{s}}\right) \\
& \times \left(\dfrac{T}{10^8~\mathrm{K}}\right)^{-2}\left(\dfrac{R}{10^6~\mathrm{cm}}\right)^{-2}~,
\end{align}
and the Ekman timescale as

\begin{align}
\nonumber \tau_{\rm Ek} \sim 10^3~\mathrm{s} &\left(\dfrac{x_n}{0.05}\right)^{-3/4}  \left(\dfrac{\rho_n}{10^{14}~\mathrm{g~cm^{-3}}}\right)^{-1/4}  \left(\dfrac{P_{\mathrm{rot}}}{0.1~\mathrm{s}}\right)^{1/2} \\
& \times \left(\dfrac{T}{10^8~\mathrm{K}}\right)\left(\dfrac{R}{10^6~\mathrm{cm}}\right)~.
\end{align}

\section{Experiment and numerical methods} 
\label{sec:exp_methods}

\begin{figure}
    \centering
    \includegraphics[width=\columnwidth]{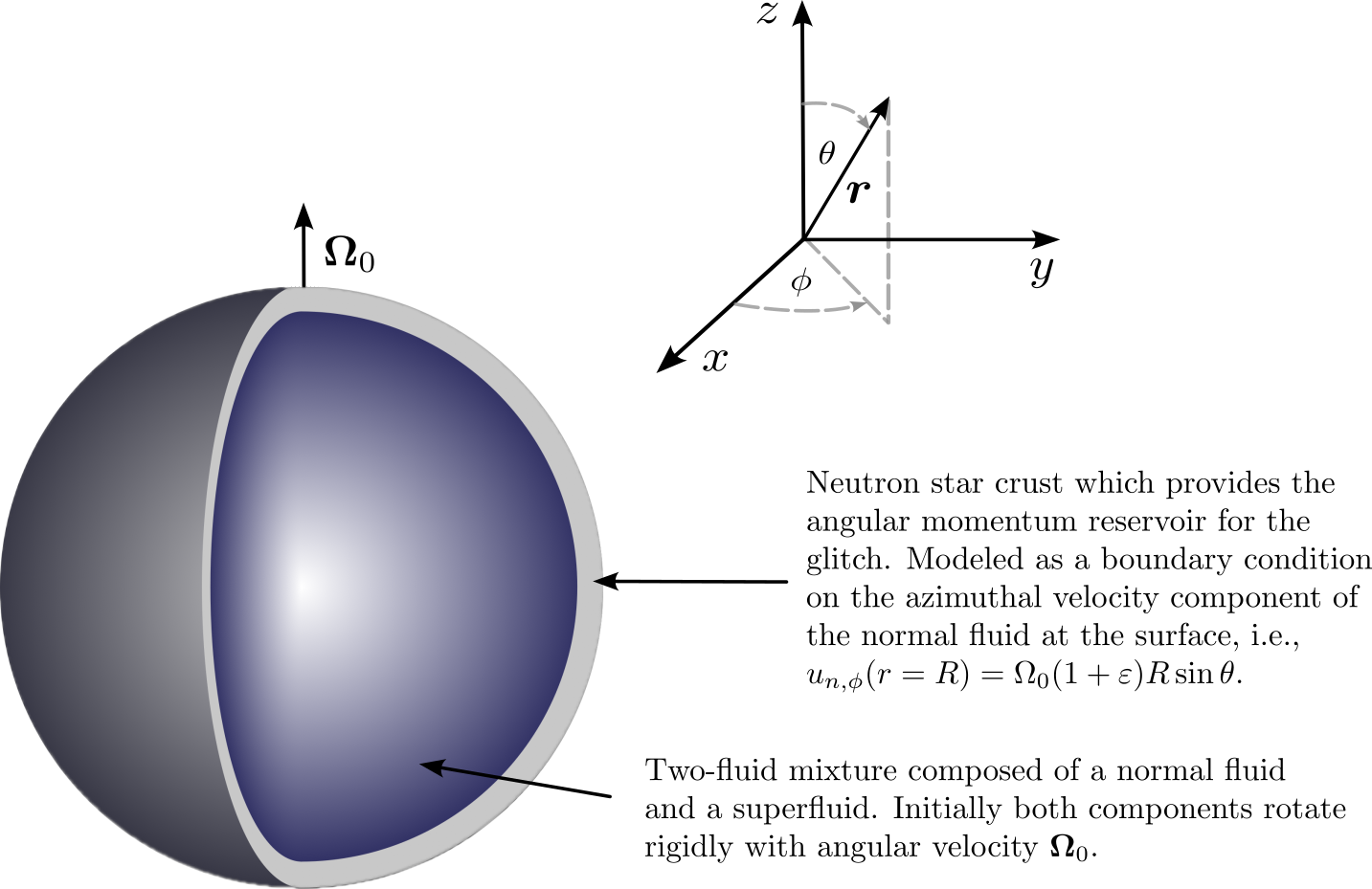}
    \caption{Illustration of our simplified model for the spin-up of a two-component fluid following a glitch. As discussed in the text, both components coexist everywhere inside the sphere, and initially both rotate with the same angular frequency $\Omega_0$ around the $z$-axis. At $t_0$, we force the azimuthal velocity of the normal component to follow $\Omega_0(1+\varepsilon)R\sin\theta$, i.e., the rotation frequency increases by a factor $\varepsilon$ with respect to the initial value.}
    \label{fig:scheme}
\end{figure}

We use hydrodynamical simulations to investigate the mutual friction coupling in a two-component fluid, and the global flow pattern after the spin-up due to a glitch. In doing so, we model the following situation: Imagine a sphere containing a neutron superfluid and a normal fluid of protons and electrons. Both components initially rotate with the same angular frequency $\Omega_0$ around the vertical axis. Then, we suddenly increase the rotation rate of the normal component at the surface of the sphere to $\Omega_0(1 + \varepsilon)$, where $\varepsilon \ll 1$ is the change in the rotation frequency (see Figure~\ref{fig:scheme}). We set out to answer the question: How does the neutron superfluid respond to such sudden acceleration, i.e., to a glitch initiated in the solid crust of a neutron star?

Unlike previous works that have investigated similar problems on spherical shells \citep[e.g.,][]{Peralta2005,Peralta_2006,Peralta2008,Howitt2016}, we adopt the entire spherical domain (i.e., including the singularity at $r=0)$ for the first time. We nondimensionalize the system of equations presented in Section~\ref{sec:hydro} using the initial rotation frequency $\Omega_0^{-1}$ and the star's radius $R$, as units of time and length, respectively. Under this choice, the evolution of the flow is characterized by the Ekman number $\mathrm{Ek}$, the mass fraction of each component ($x_s = \rho_s/\rho$ and $x_n=\rho_n/\rho)$, and the coefficients of mutual friction $\mathcal{B}$ and $\mathcal{B}'$. 

In the bulk of a typical neutron star, the Ekman number is very small (i.e., $\mathrm{Ek}\sim 10^{-9}$; see Equation~\eqref{eq:ek_n}). Consequently, viscous boundary layers with a characteristic scale of $\delta_\nu \sim R\sqrt{\rm Ek}$ are very small (i.e., of order $\delta_\nu\sim 30\, {\rm cm}$ for $R\sim 10 \, {\rm km}$), demanding high spatial resolutions to fully resolve them. As this is beyond the limitations of current computational capabilities, we restrict our study to $\mathrm{Ek}\sim 10^{-4}$--$10^{-3}$. We also take $x_s = 0.95$ and $x_n = 0.05$, which are similar to values expected for neutron star interiors and superfluid helium \citep[see, e.g.,][]{Barenghi1983, Glampedakis2011, Graber_2017}. We further adopt $\mathcal{B}/\mathcal{B}' \approx 2$, with $\mathcal{B}$ ranging from $10^{-3}$ to 1, in order to explore the regime of weak mutual friction (most likely relevant for neutron stars; see Section~\ref{sec:mf}), and strong mutual friction \citep[relevant for laboratory experiments with superfluid helium;][]{Tsakadze_and_Tsakadze_1980, Barenghi1983}. At this point we also highlight that we do not find differences in our results when changing the ratio $\mathcal{B}/\mathcal{B}'$ or neglecting $\mathcal{B}'$, which suggests that $\mathcal{B}'$ does not play a dominant role in the spin-up dynamics \citep[as also found by][]{van_Eysden_2013}. Finally, we use a relative glitch size of $\varepsilon = 10^{-3}$. While larger than observed glitches \citep[e.g.,][]{Espinoza_2011,Basu2022}, we opt for this value to reduce computation times of steady flow patterns.

The boundary conditions satisfied by the normal fluid component at the star's surface ($r=R=1$) are impenetrable and no-slip, i.e., $u_{n,r} = u_{n,\theta} = 0$, and $u_{n,\phi} = \Omega_0 (1+\varepsilon)R\sin\theta$.  For the neutron superfluid, the boundary conditions depend on the interaction between vortex lines and the surface \citep[see, e.g.,][for an extensive discussion on this topic]{Peralta2005}. For simplicity, we impose impenetrable and stress-free boundary conditions ($u_{s,r} = \partial_r(u_{s,\theta}/r) =  \partial_r(u_{s,\phi}/r)=0$).

We solve Equations~\eqref{eq:u_n}--\eqref{eq:mass_conservation}  along with the boundary conditions using the pseudo-spectral solver Dedalus \citep{Burns2020}. The variables are represented in spherical harmonics for the angular directions and Jacobi polynomials for the radial direction. All our simulations have $L_{\mathrm{max}} = N_{\mathrm{max}} = 255$, where $L_{\mathrm{max}}$ is the maximum spherical harmonic degree, and $N_{\mathrm{max}}$ is the maximal degree of the Jacobi polynomials used in the radial expansion. Therefore, the number of radial, latitudinal, and longitudinal points are $(N_r,N_\theta,N_\phi) = (256,256,512)$, respectively. For time-stepping, we use a second order semi-implicit BDF scheme \citep[SBDF2,][]{wang_ruuth_2008}, where the linear and nonlinear terms are treated implicitly and explicitly, respectively. To ensure numerical stability, the size of the time steps is set by the Courant–Friedrichs–Lewy (CFL) condition, using a safety factor of 0.2 (based on trial an error). To prevent aliasing errors, we apply the ``3/2 rule'' in all directions when evaluating nonlinear terms.

\section{Analysis and Results} 
\label{sec:results}

Since our simulations, to the best of our knowledge, are the first to treat the spin-up problem in the full sphere, we first investigate the spin-up of a purely viscous (normal) flow due to Ekman pumping in Section~\ref{sec:spin_up_n}. Then, in Section~\ref{sec:spin_up_s} we investigate the spin-up of a two-component fluid in detail. In what follows, all numerical results are presented in dimensionless form.

\begin{figure}
    \centering
    \includegraphics[width=\columnwidth]{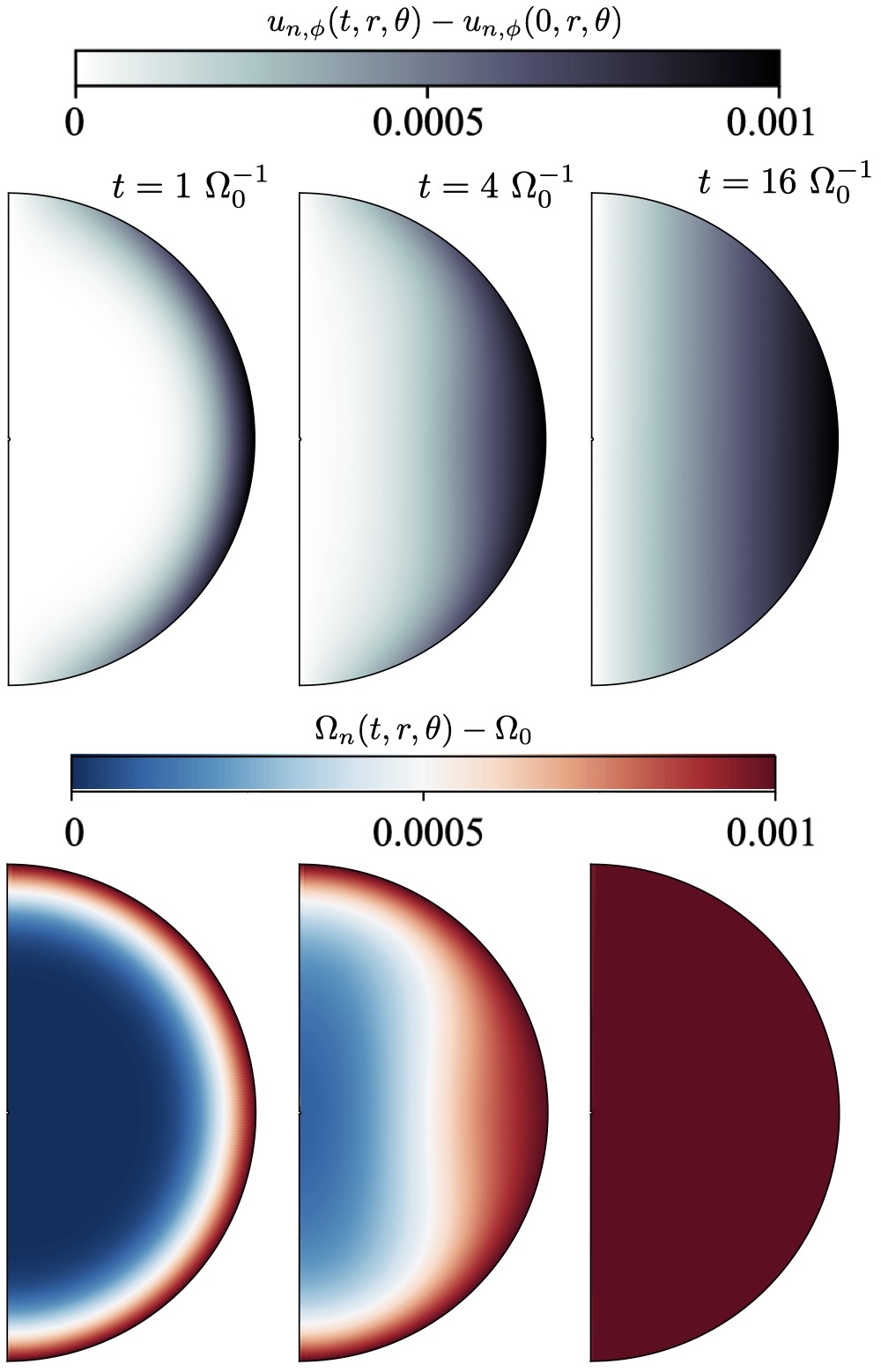}
    \caption{Azimuthal component of the velocity and rotation frequency (both with respect to the initial state) at different times after the sudden change in the boundary's rotation rate (top and bottom panels, respectively). The results show the spin-up of a purely viscous fluid with Ekman number $\mathrm{Ek} = 5\times 10^{-3}$. As expected, the spin-up occurs due to Ekman pumping and angular momentum is distributed from the equator to the interior, with the azimuthal velocity being constant on cylindrical surfaces at fix radius.}
    \label{fig:ekman_1}
\end{figure}

\subsection{Spin-up of a purely viscous fluid} 
\label{sec:spin_up_n}

\begin{figure*}
    \centering
    \includegraphics[width=\textwidth]{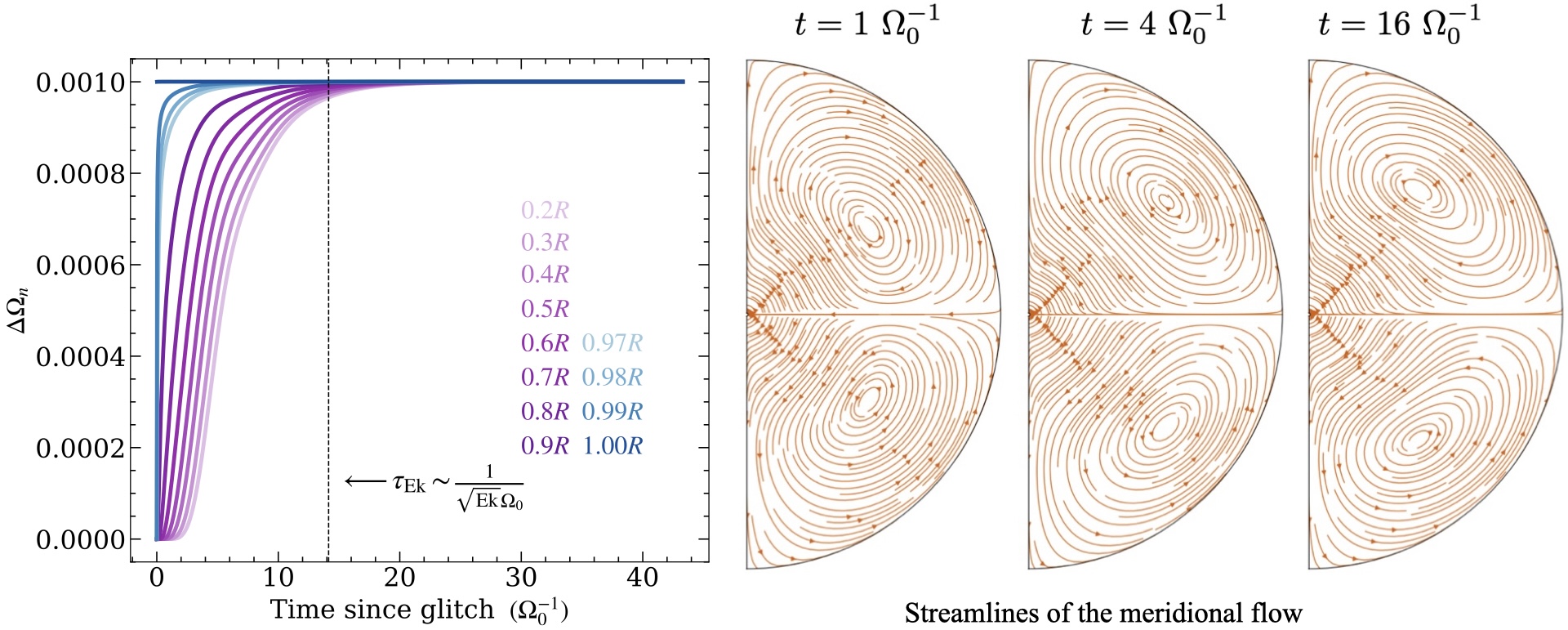}
    \caption{Left panel: Change in the rotation frequency of a viscous fluid, $\Delta \Omega_n = \Omega_n - \Omega_0$, as a function of time since the imposed change at the boundary. $\Omega_n$ is measured at the equator and at different radial locations (distinguished by different colors as shown in the plot). The results presented in this figure correspond to a run using $\mathrm{Ek} = 5\times 10^{-3}$, where the Ekman timescales is $\tau_{\mathrm{Ek}}\approx 14~\Omega_0^{-1}$ (dashed vertical line). At later times, the fluid reaches solid body rotation with $\Delta\Omega_n \approx \varepsilon\Omega_0 = 10^{-3}$. We do not show results for $r$ closer to 0 because small fluctuations in the azimuthal flow velocity introduce plotting artifacts when calculating the rotation frequency via $\Omega_n = u_{n,\phi}/(r\sin\theta)$. Middle and right panels: Meridional streamlines at different times. The single-cell patterns in the northern and southern hemisphere circulate in opposite directions and persist across the simulation.}
    \label{fig:ekman_2}
\end{figure*}

Once the surface of the sphere has been suddenly accelerated, a viscous boundary layer forms, wherein the flow rotates faster than in the interior of the fluid (far away from the boundary). For our experiment with a viscous single-component fluid, we set $\mathrm{Ek} = 5\times 10^{-3}$ which corresponds to a boundary layer of thickness $\delta_\nu \sim 0.07R$. This layer acts as a source of angular momentum and accelerates the bulk of the fluid by the process of Ekman pumping outlined previously. 

Figure~\ref{fig:ekman_1} highlights the evolution of the azimuthal flow (top panels), and the rotation frequency $\Omega_n(t,r,\theta) = u_{n,\phi}(t,r,\theta)/(r\sin\theta)$ (bottom panels), following the initial acceleration of the boundary. Both quantities are displayed at different times, relative to the initial corotation state. We find that the angular momentum is first distributed to the equatorial region and then re-circulated inwards. As the evolution progresses, the azimuthal velocity becomes axisymmetric with constant magnitude over cylindrical surfaces as a direct consequence of the Taylor--Proudman theorem \citep{Proudman1916,Taylor1917}. As a result, we find $u_{n,\phi} \rightarrow \Omega_0(1+\varepsilon) r\sin\theta$, i.e., a state of solid-body rotation with $\Omega_n \approx \Omega_0(1+\varepsilon)$.

As discussed in Section~\ref{sec:mf}, the spin-up timescale of a viscous flow due to Ekman pumping is $\tau_{\mathrm{Ek}} \sim 1/(\sqrt{\mathrm{Ek}}\Omega_0)$, which is $\approx 14~\Omega_0^{-1}$ in our example. Because the fluid accelerates first at the equator and the velocity then becomes constant over cylindrical surfaces, we can follow the spin-up behaviour in time by calculating the rotation frequency at fixed colatitude $\theta = \pi/2$ (the equator) for different radial locations. Our results are shown in the left panel of Figure~\ref{fig:ekman_2}. We find that close to the surface (blue curves), the flow accelerates on a timescale that is much shorter than that for the interior layers (purple curves). However, as expected, the bulk of the fluid spins up to the new angular velocity within an Ekman timescale (dashed vertical line). Finally, the meridional streamlines of the viscous flow exhibit a single cell in each semihemisphere (symmetric with respect to the equator but circulating in opposite directions), remaining similar at all times (see middle and right panels of Figure~\ref{fig:ekman_2}).

\subsection{Spin-up of a two-component fluid} 
\label{sec:spin_up_s}

The spin-up of a two-component fluid, where one component is superfluid, differs from that of a purely viscous fluid because the normal (viscous) fluid and superfluid components coexist and interact with each other. Yet, the initial evolution of the system is qualitatively similar: Once the normal component of the flow is suddenly accelerated at the surface, the interior layers spin up due to Ekman pumping. As soon as differential rotation builds between the two fluids, the superfluid component couples to the normal flow due to mutual friction, accelerating on a much longer timescale that depends on the magnitude of the friction coefficient $\mathcal{B}$ and the Ekman number $\mathrm{Ek}$. In all our simulations, the system reaches a steady state where the normal component and the superfluid co-rotate with the same angular frequency across the entire spherical domain.

\begin{figure}
    \centering
    \includegraphics[width=\columnwidth]{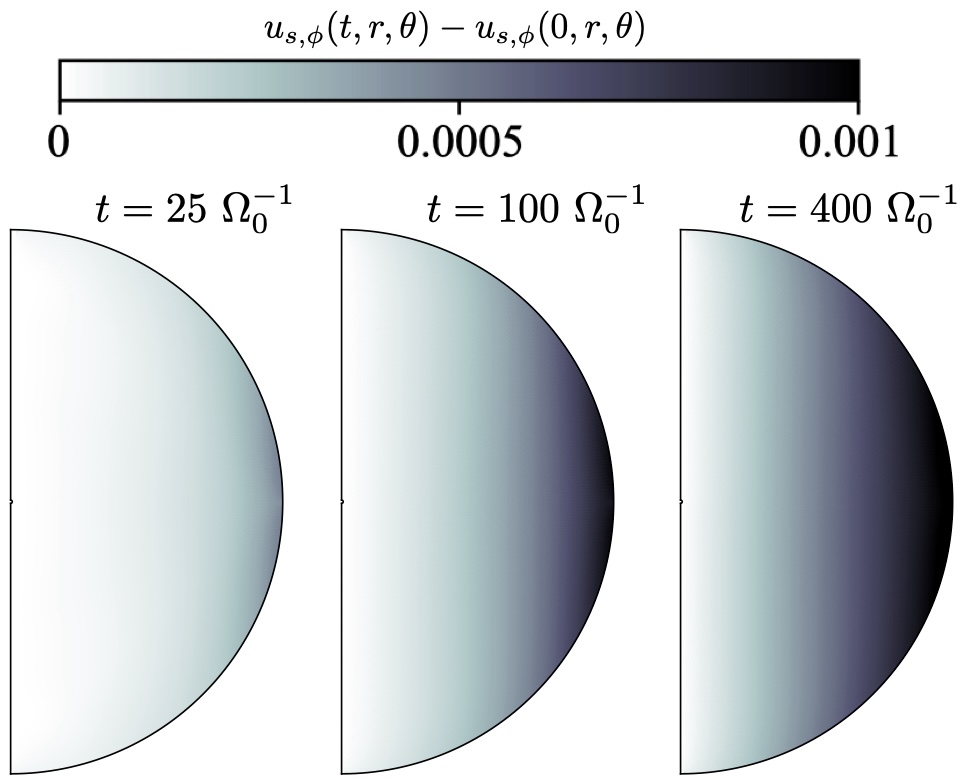}
    \caption{Azimuthal component of the superfluid velocity with respect to the initial state at different times during the spin-up. Results are shown for the run using $\mathrm{Ek} = 5\times 10^{-3}$ and $\mathcal{B} \approx 0.01$. The spin-up occurs on longer timescales compared to the viscous case owing to mutual friction. Angular momentum is distributed from the equator to the interior, with the azimuthal velocity being constant on cylindrical surfaces at fix radii.}
    \label{fig:azimuthal_flow}
\end{figure}

\begin{figure}
    \centering
    \includegraphics[width=\columnwidth]{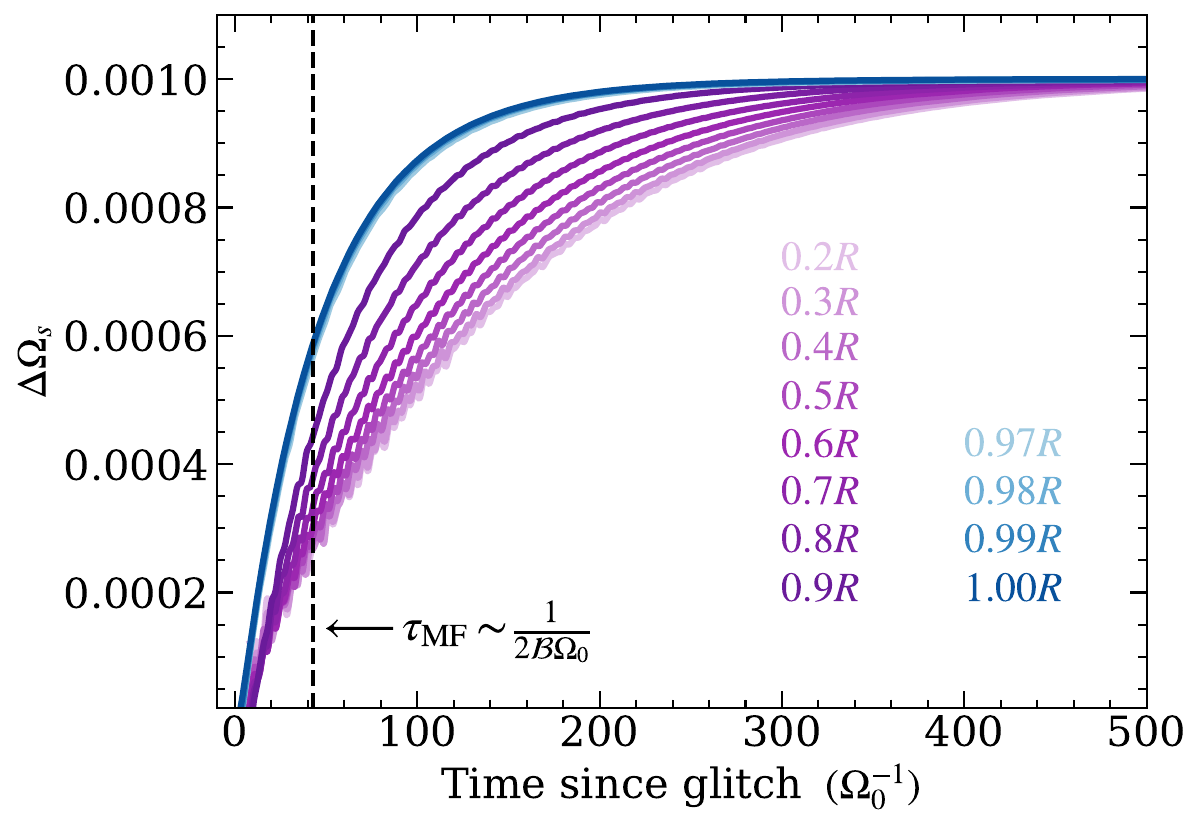}
    \caption{Change in the rotation frequency of the superfluid component, $\Delta \Omega_s = \Omega_s - \Omega_0$, as a function of time since the imposed acceleration of the boundary. $\Omega_s$ is measured at the equator and at different radial locations (distinguished by different colors as highlighted in the plot). Results are shown for the run using $\mathrm{Ek} = 5\times 10^{-3}$ and $\mathcal{B} \approx 0.01$, where the Ekman and mutual friction timescales are $\tau_{\mathrm{Ek}}\approx 14~\Omega_0^{-1}$ and $\tau_{\mathrm{MF}}\approx 43~\Omega_0^{-1}$ (dashed vertical line), respectively. At later times, the fluid reaches solid body rotation with $\Delta\Omega_s \approx \varepsilon\Omega_0 = 10^{-3}$.}
    \label{fig:time_series}
\end{figure}

Figure~\ref{fig:azimuthal_flow} shows the evolution of the azimuthal superfluid flow (relative to the initial state) at different times after the initial acceleration of the boundary. We only display the superfluid and exclude the normal component since it evolves in a qualitatively similar way to the purely viscous case discussed in Section~\ref{sec:spin_up_n}. As for spin-up via Ekman pumping in a viscous flow, the superfluid's azimuthal velocity becomes axisymmetric, reaching a state of solid-body rotation with $\Omega_s \approx \Omega_0(1+\varepsilon)$. Time series of the angular frequency at different radial locations show that close to the surface (blue curves), the superfluid accelerates on a timescale that is consistent with mutual friction, while it takes longer for the interior layers (purple curves) to spin up (see Figure~\ref{fig:time_series}). We note that near the center, we observe small oscillations in the angular frequency, which may result from the more complex flow pattern near $r=0$, and interactions between the superfluid and the normal component. Nonetheless, the oscillations fade away over time and do not affect our measurements of the spin-up time (see below) of the innermost layers.

\begin{figure}
    \centering
\includegraphics[width=\columnwidth]{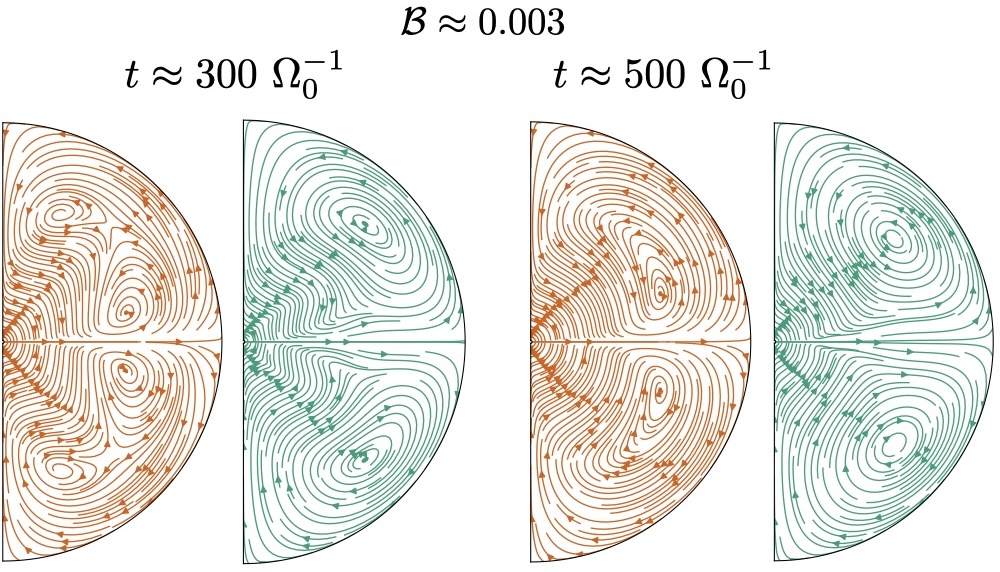}
\includegraphics[width=\columnwidth]{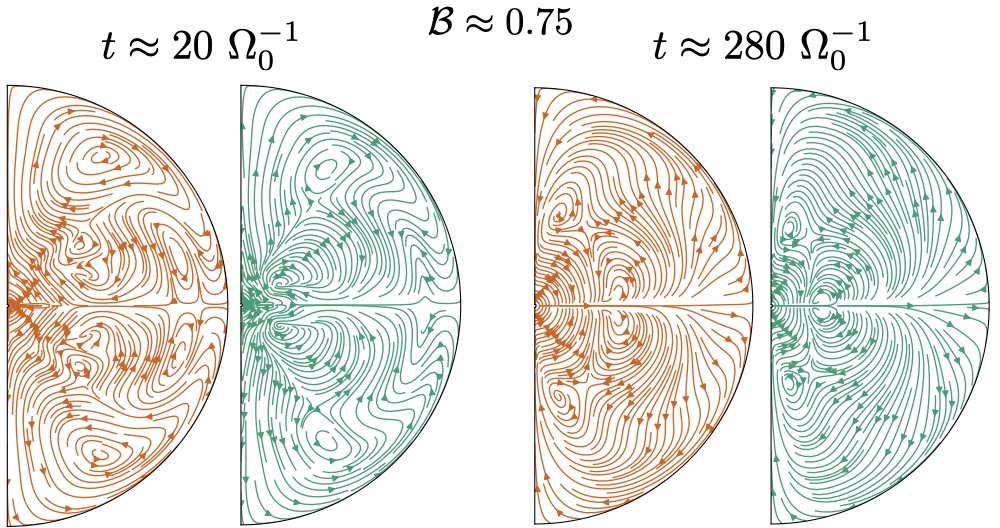}
    \caption{Meridional streamlines for the normal (orange) and superfluid (green) components. Results are shown at two particular times, for runs at $\mathrm{Ek} = 5\times 10^{-3}$ using $\mathcal{B} \approx 0.003$ and $\approx 0.75$, to illustrate how the strength of coupling between the two component influences the flow pattern. Note that the streamlines are similar for the strong coupling regime (bottom panels), and different for the weak coupling regime (upper panels).}
    \label{fig:B_streamlines}
\end{figure}

Figure~\ref{fig:B_streamlines} displays poloidal streamlines for two special cases using $\mathrm{Ek} = 5\times 10^{-3}$, where $\mathcal{B}\approx 0.003$ (weak coupling, upper panels) and $\mathcal{B}\approx 0.75$ (strong coupling, lower panels). We first note that in the weak coupling regime, the flow pattern of the superfluid and normal component seem to evolve independently from each other (due to the small mutual friction). However, in the strong coupling regime, the superfluid is dragged along with the normal component, so that both flow patterns exhibit similar structures at all times. In general, the meridional circulation of the system is complex and resembles structures reported in previous studies using spherical shells \citep[e.g.,][]{Peralta2005,Peralta_2006,Peralta2008}. We also observe that unlike for the spin-up of the purely viscous fluid, where a single meridional circulation cell persists per semihemisphere at all times (see Figure~\ref{fig:ekman_2}), the interaction between the normal and superfluid components causes the flow pattern to contain multiple circulation cells with smaller Taylor vortices \citep{Hollerbach1998} emerging and disappearing intermittently, particularly at early times after the acceleration of the boundary. Once both components of the fluid achieve solid-body rotation, the profile of the meridional circulation becomes steady in time. However, only in the weak coupling regime does the flow pattern transition to a single cell, similar to the purely viscous case.

We are interested in characterizing the spin-up timescale of the superfluid, $t_{\rm{spin-up}}$, in the interior of the star. Since the fluid rotates with constant angular frequency over cylindrical surfaces in radius, we focus on the equatorial plane at different radial locations. Then, to measure $t_{\rm{spin-up}}$, we create time series of $\Delta \Omega_s$ (such as the ones displayed in Figure~\ref{fig:time_series}), and fit a function of the form $\Delta \Omega_s = \varepsilon(1 - e^{-t/t_{\rm{spin-up}}})$ to extract the timescale. 

\begin{figure*}
    \centering
    \includegraphics[width=\textwidth]{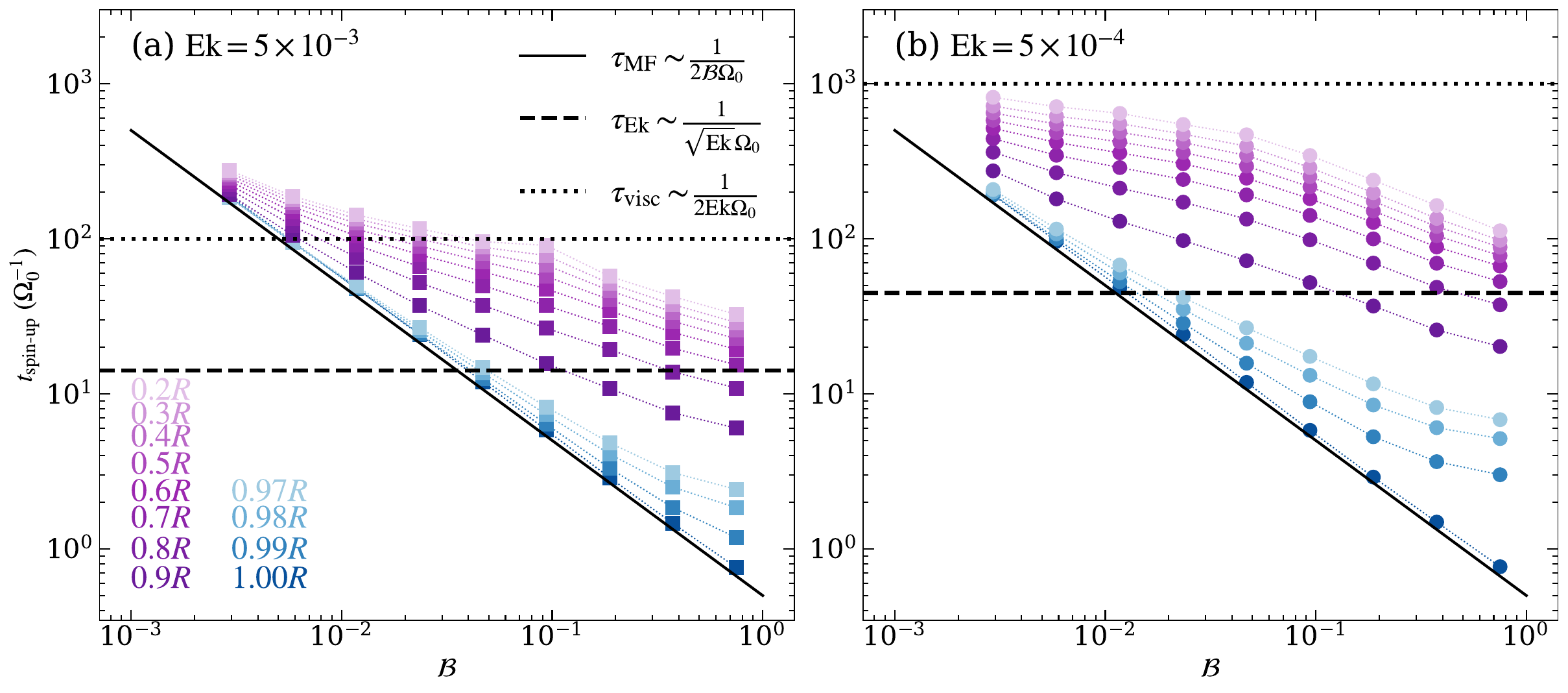}
    \caption{Spin-up timescale, $t_{\rm{spin-up}}$, of the superfluid component as a function of the mutual friction coefficient $\mathcal{B}$. Panel (a) and (b) show results for simulations at $\mathrm{Ek} = 5\times 10^{-3}$ and $\mathrm{Ek} = 5\times 10^{-4}$, respectively. Colors distinguish between measurements at different radial locations in the interior flow ($0.2R$ -- $0.9R$ in a purple scale, and locations within the viscous boundary layer of the normal component; $0.97R$ -- $R$, in a blue scale). We also show the spin-up timescale of the superfluid expected from mutual friction (solid line) and the Ekman and viscous spin-up timescales of the normal component (dashed and dotted lines). We do not show results for $r$ closer to 0 because small fluctuations in the azimuthal flow velocity introduce plotting artifacts when calculating the rotation frequency $\Omega_s = u_{s,\phi}/(r\sin\theta)$ (see text).}
    \label{fig:spin-up_t_scale}
\end{figure*}

Figure~\ref{fig:spin-up_t_scale} shows the results for all our simulations. At the surface ($r=R$), we find that regardless of the value of $\mathcal{B}$ and $\mathrm{Ek}$, there is an excellent agreement between our measurements and the timescale $\tau_{\rm MF}$ expected from mutual friction. Close to the surface, within the viscous boundary layer of the normal component ($r \approx 0.97R$--$R$, blue markers), the mutual friction timescale continues to capture the spin-up behaviour well for low values of $\mathcal{B}$, while small deviations are observed for $\mathcal{B} \gtrsim 0.1$. In general, deviations from $\tau_{\rm MF}$ become significant in the interior regions, particularly when the coupling strength between the normal and superfluid components increases. This is expected since at larger $\mathcal{B}$, the dynamics of the superfluid become strongly influenced by the normal component. Note that for the regions far from the boundary ($r \leq 0.9R)$ there seems to be a change in the slope of the curves around $\mathcal{B} \approx 0.1$. As we discuss later, this could be related to a change in the coupling regime. 

Comparing the left and right panels in Figure~\ref{fig:spin-up_t_scale} constructed from runs with two different Ekman numbers ($\mathrm{Ek} = 5\times 10^{-3}$ and $\mathrm{Ek} = 5\times 10^{-4}$, respectively), we also find that the spin-up timescale of the interior layers of the superfluid becomes longer for smaller $\mathrm{Ek}$. Consequently, the deviations from $\tau_{\rm MF}$ become more noticeable. This makes sense because the smaller the viscosity (or the smaller the Ekman number), the longer it takes for Ekman pumping to occur, delaying first the spin-up of the interior layers of the normal fluid, and subsequently, the spin-up of the bulk of the superfluid.

These results suggest that $t_{\rm{spin-up}}$ for the interior layers ($r\leq 0.9R$) is not solely a function of $\mathcal{B}$. Instead, the timescale should also depend on the Ekman number (Ek) and/or the radial position $r$ within the sphere. In particular, the fact that the spin-up time of the interior layers in the simulations shows a similar slope as a function of $\mathcal{B}$ for both values of Ek (as discussed above this is true for $\mathcal{B} \gtrsim 0.1$ but not for smaller values of $\mathcal{B}$), suggests a power-law relationship of the form

\begin{align}\label{eq:spinup_r}
t_{\rm{spin-up}}(r) \sim C \tau^\alpha_{\mathrm{Ek}}[L(r)] \mathcal{B}^{-\gamma}
    \approx C \tau^\alpha_{\mathrm{Ek}}(\Delta R) \mathcal{B}^{-\gamma} ~,
\end{align}
where $C$, $\alpha$, and $\gamma$ are constants, and we estimate the relevant length-scale $L(r)$ for calculating the Ekman time for a fluid layer at $r$ as $L(r) \approx \Delta R = R -r$.

\begin{figure}
    \centering
    \includegraphics[width=\columnwidth]{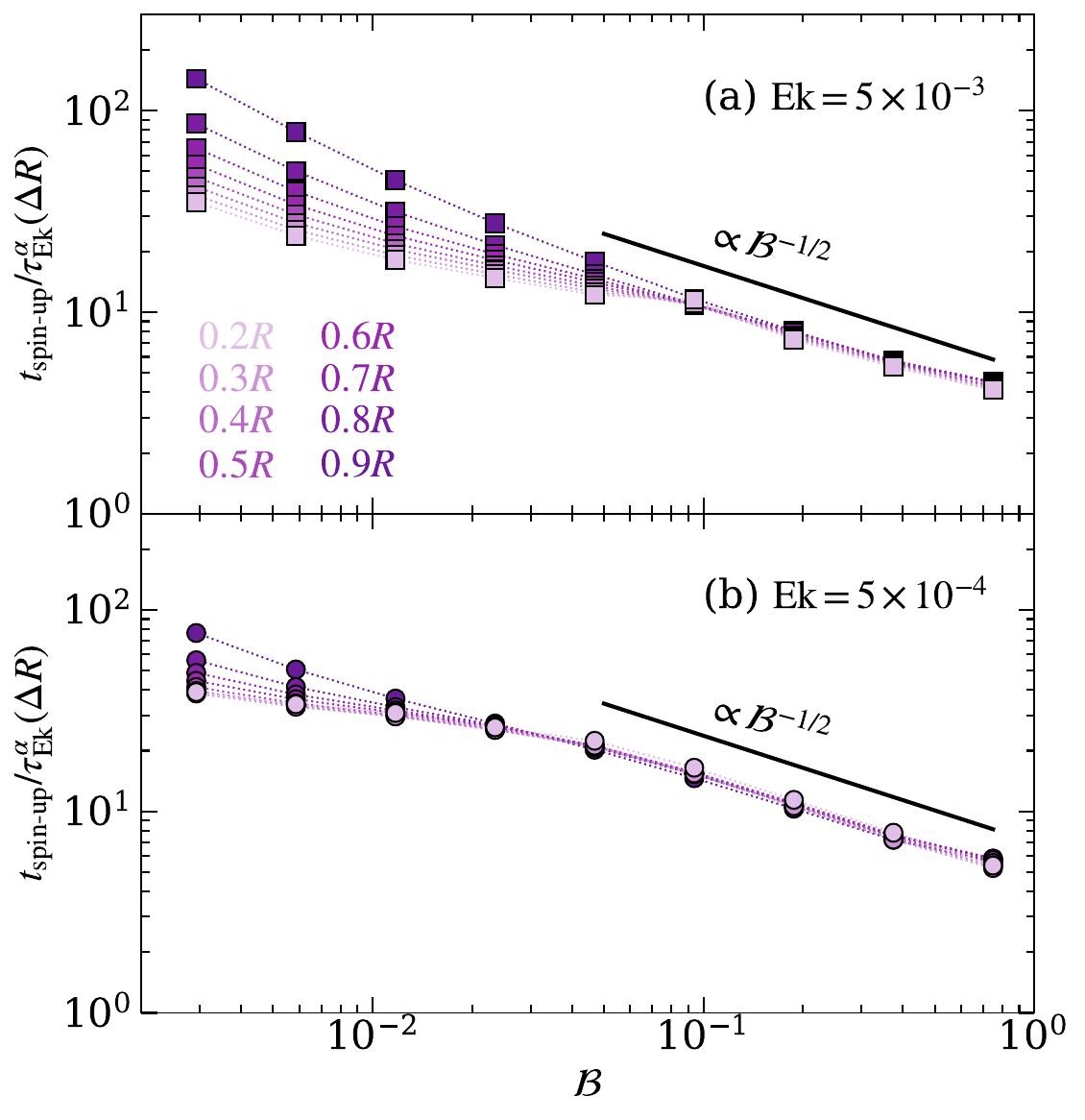}
    \caption{Ratio between $t_{\rm{spin-up}}$ and $\tau^{\alpha}_\mathrm{Ek}(\Delta R)$ (where $\Delta R = R - r$ corresponds to the characteristic lengthscale used to determine the Ekman time at a given $r$), as a function of the mutual friction coefficient $\mathcal{B}$. Panel (a) and (b) show results for simulations at $\mathrm{Ek} = 5\times 10^{-3}$ and $\mathrm{Ek} = 5\times 10^{-4}$, respectively. As discussed in the text, $\alpha = 0.85$ allows to superpose the measurements of the spin-up timescales at different radial locations $r$ for $\mathcal{B} \gtrsim 0.1$. The black solid lines are straight lines of slope $-1/2$ in log-log space to illustrate the dependence on $\mathcal{B}$ for $\mathcal{B} \gtrsim 0.1$.}
    \label{fig:spin-up_t_scale_norm}
\end{figure}

When fitting Equation~\eqref{eq:spinup_r} to our measured spin-up timescales, we find that $\alpha \approx 0.85$ and $\gamma \approx 0.5$ are the best parameters to describe the data when $\mathcal{B} \gtrsim 0.1$ (the strong coupling regime), regardless of the value of the Ekman number (see Figure~\ref{fig:spin-up_t_scale_norm}). This dependence with the Ekman time provides reasonable fits even for values down to $\mathcal{B} \approx 0.01$, but only for the small Ek runs (bottom panel of Figure~\ref{fig:spin-up_t_scale_norm}). The proportionality constant is $C = 3.2$ for $\mathrm{Ek} = 5\times 10^{-3}$, and $C=4.6$ for $\mathrm{Ek} = 5\times 10^{-4}$. The characterization of the timescale for $\mathcal{B} \lesssim 0.1$ is complicated due to the large dispersion in the measurements. A possible reason for this dispersive behavior could be that weaker coupling between the superfluid and the normal component weakens the dependence of $t_{\rm{spin-up}}$ with Ek, such that a function of the form \eqref{eq:spinup_r} is no longer valid in this regime. This, together with the fact that $\tau_{\mathrm{Ek}}\propto \Delta R$, with $\Delta R$ being smaller closer to the surface, causes the ratio between the two timescales to increase as we move towards the surface. Moreover, the dispersion for low $\mathcal{B}$ values is also smaller for the smaller Ek case. This could be due to the fact that transport of angular momentum by viscous diffusion and not Ekman pumping might be an additional factor that modifies the dynamical coupling timescales of the two-fluid system. This scenario could be tested with additional simulations at even smaller values of Ek to see whether the dispersion decreases further. However, such explorations were beyond the scope of this paper and we leave this for future work.

\section{Discussion} 
\label{sec:discussion}

\subsection{Summary of our results}

In this work, we have studied the spin-up dynamics of a two-component system, composed of a superfluid and a normal (viscous) fluid, in the context of neutron star glitches. Unlike previous work which was limited to conducting simulations of spherical shells—imposing artificial boundary conditions on the inner surface and potentially altering the dynamics—we solve the non-linear HVBK equations for such a superfluid system for the entire sphere, including $r=0$, for the first time. We confirm that in the absence of the superfluid component, the spin-up of a purely viscous fluid occurs due to the phenomenon of Ekman pumping, and the bulk of the fluid responds to the sudden acceleration of the surface of the sphere on a timescale $\tau_{\mathrm{Ek}}\sim 1/(\sqrt{\mathrm{Ek}}\Omega_{0})$ (see left panel in Figure~\ref{fig:ekman_2}). The resulting meridional flow is characterized by a single cell in each semihemisphere of the sphere which persists at all times (see middle and right panels in Figure~\ref{fig:ekman_2}). 

The spin-up of a two-component fluid is more complex and involves different timescales. The normal component of the fluid spins up due to Ekman pumping, but only the surface layers accelerate on an Ekman time $\tau_{\mathrm{Ek}}$. The interior flow responds on a time span that is longer than $\tau_{\mathrm{Ek}}$. This is because the mutual friction force, which couples the normal and the superfluid component, drives an exchange of angular momentum from the normal component to the superfluid component. This consequently delays the response of the internal normal-fluid layers, while also causing the superfluid to spin up on multiple timescales (Figure~\ref{fig:time_series}). The surface layers of the superfluid accelerate on a mutual friction timescale $\tau_{\mathrm{MF}}\sim 1/(2\mathcal{B}\Omega_0)$, while the interior superfluid layers spin up on a much longer timescale, depending on the strength of the coupling coefficient $\mathcal{B}$ and the Ekman number (Figure~\ref{fig:spin-up_t_scale}). Unlike the purely viscous case, the resulting meridional flow patterns of both components are complex and characterized by multiple cells that appear and disappear as the system evolves (Figure~\ref{fig:B_streamlines}). However, once the system reaches a steady state, the flow pattern becomes more similar to the purely viscous case, particularly in the weak coupling regime.

When measuring the spin-up timescale of the interior layers of the superfluid, we have further found two different spin-up regimes, depending on the value of $\mathcal{B}$ (Figure~\ref{fig:spin-up_t_scale_norm}). When the coupling between the normal and the superfluid component is strong ($\mathcal{B}\gtrsim 0.1$), the spin-up time of the interior flow at a radial location $r$ follows $t_{\mathrm{spin-up}}(r) \propto \tau_{\mathrm{Ek}}^{0.85}(\Delta R)~\mathcal{B}^{-1/2}$, where $\tau_{\mathrm{Ek}}(\Delta R)$ is the Ekman time for a layer of thickness $\Delta R = R - r$ (see Equation~\ref{eq:spinup_r}). On the other hand, when the coupling between both components is weak ($\mathcal{B}\lesssim 0.1$), we have found no clear relation between the spin-up timescale of the interior layers as a function of radius, Ek, or $\mathcal{B}$. However, it is worth noting that deviations from Equation~\eqref{eq:spinup_r} in the weak coupling regime are much smaller for simulations using the lower Ekman number $\mathrm{Ek} = 5\times 10^{-4}$ (compare the two panels of Figure~\ref{fig:spin-up_t_scale_norm} at low values of $\mathcal{B}$).

\subsection{Comparison to previous work}

Comparison with previous work is difficult because, to our knowledge, the specific problem discussed in this paper has not been previously investigated using numerical simulations. Similar studies have focused on instabilities and flow patterns in superfluid Couette flow within spherical shells, as explored in a series of papers by \cite{Peralta2005,Peralta_2006,Peralta2008}. The profiles of the meridional circulation observed in our simulations resemble those reported in these studies, where multiple cells and Taylor vortices appear in both the normal and superfluid components of the flow. While previous work showed these structures to persist over time, likely due to a non-zero mutual friction force sustained by an imposed differential rotation, our simulations evolve towards a steady state of solid-body rotation. Consequently, the complex flow patterns observed initially disappear when the system is evolved long-term.

The spin-up of a two-component fluid in the full sphere was also investigated analytically by \citet{van_Eysden_2013} neglecting the non-linear terms of the HVBK equations. Interestingly, they also observed a regime separation depending on the strength of the mutual friction coupling between the normal and superfluid components. When $\mathcal{B} \sim 1$, the superfluid and normal components are strongly coupled, and any differential velocity between the components is removed by mutual friction on a short timescale, so that both components are locked together and accelerate on an Ekman pumping timescale. Alternatively, for $\mathcal{B} \sim \sqrt{\mathrm{Ek}} \ll 1$, \citet{van_Eysden_2013} find no Ekman pumping in the superfluid component and the system spins up via mutual friction on a complex combination of the mutual friction and the Ekman timescales (although not explicitly given in their work). However, the measured timescales in our simulations differ from the predictions in \cite{van_Eysden_2013}. For instance, even in the strong coupling regime, the superfluid spins up over a timescale that is longer than the Ekman timescale and aligns well with Equation~\eqref{eq:spinup_r}. These discrepancies are not surprising because their analytic calculations neglect the non-linear dynamics of the system. Nonetheless, it is encouraging that our simulations agree, at least partially, with their results. For example, the regime separation at $\mathcal{B} \sim \sqrt{\mathrm{Ek}}$ corresponds to $\mathcal{B} \approx 0.07$ (0.02) for $\mathrm{Ek} = 5\times 10^{-3} (5\times 10^{-4}$) in our simulations. When comparing with Figure~\ref{fig:spin-up_t_scale_norm}, these values of $\mathcal{B}$ accurately delimit the validity of Equation~\eqref{eq:spinup_r}, and the beginning of the large dispersion in the measured spin-up timescales of the interior layers appearing in the weaker coupling regime.

\subsection{Limitations and future work}

While our choices for relevant numerical parameters were motivated by neutron star physics, the assumptions and approximations in our model and simulations limit the direct applicability of our results to real neutron stars. For instance, we have neglected density stratification and magnetic fields. Spin-up in a stratified fluid is different because buoyancy may inhibit the radial motion of fluid elements, limiting the extent of secondary circulations such as the Ekman flow \citep[e.g.,][]{Clark1971,Clark1973}. While the HVBK equations can be easily generalized to a stratified and fully-compressible two-component fluid, numerical solutions are computationally challenging. This difficulty arises because the compression and expansion of the fluid generate sound waves, which usually propagate much faster than typical fluid flows inside neutron stars. As a result, much smaller computational time steps are needed to resolve these waves in numerical simulations. Moreover, taking into account magnetic fields would require adding the Lorentz force to the momentum equation of the normal component as well as solving the magnetic induction equation. As a result, the coupling between the surface and interior flows could occur on shorter timescales \citep[e.g.,][]{Mendell2001}.

In addition, the Ekman numbers considered in this work are significantly larger compared to real values in neutron stars. This could introduce two complications. First, as shown in Figure~\ref{fig:spin-up_t_scale_norm}, viscous diffusion can modify the spin-up timescale in the deep interior of the star. Second, the normal component's flow in our simulations is laminar. However, turbulence can easily develop in real neutron stars \citep{Andersson2007,Haskell2020}. On one hand, turbulence can trigger differential rotation in the fluids, while it would also cause superfluid vortices to become tangled. Vortices would thus no longer remain straight (as explicitly assumed in our mutual friction force in Equation~\eqref{eq:mutual_friction}), requiring the incorporation of vortex tension and potentially more suitable forms of coupling between the superfluid and the normal component \citep[e.g.,][]{Gorter1949,Antonelli2020,Celora2020}. All these effects could give the spin-up process a completely different character to what we discussed above.

As a result, we encourage future work on the spin-up of superfluids, either in the laboratory or with numerical simulations. Improvements in numerical modeling by relaxing some of the assumptions above are needed to interpret the rich set of observations of glitches in neutron stars.

\begin{acknowledgements}
The authors thank Andrew Cumming for useful conversations and Nils Andersson, Bryn Haskell, and Toby Wood for providing valuable feedback on the manuscript. J.R.F. is supported by NASA through the Solar System Workings grant 80NSSC24K0927, and Heliophysics grants 80NSSC19K0267 and 80NSSC20K0193. 
\end{acknowledgements}

\bibliography{references}{}
\bibliographystyle{aasjournal}

\end{document}